\documentclass[aps,preprint,superscriptaddress,showpacs,amsmath,amssymb,endfloats*]{revtex4}

\usepackage{graphicx}
\usepackage{bm}

\begin{document}
\title{Increasing the coherence time of Bose-Einstein-condensate
interferometers with optical control of dynamics}
\author{James A. Stickney}
\affiliation{Department of Physics, Worcester Polytechnic Institute,
100 Institute Road, Worcester, Massachusetts 01609, USA}
\author{Dana Z. Anderson}
\affiliation{Department of Physics and JILA, University of Colorado
and National Institute of Standards and Technology, Boulder,
Colorado 80309-0440, USA}
\author{Alex A. Zozulya}
\affiliation{Department of Physics, Worcester Polytechnic Institute,
100 Institute Road, Worcester, Massachusetts 01609, USA} \email[]{
zozulya@wpi.edu}

\begin{abstract}
Atom interferometers using Bose-Einstein condensate that is confined
in a waveguide and manipulated by optical pulses have been limited
by their short coherence times.  We present a theoretical model that
offers a physically simple explanation for the loss of contrast and
propose the method for increasing the fringe contrast by recombining
the atoms at a different time. A simple, quantitatively accurate,
analytical expression for the optimized recombination time is
presented and used to place limits on the physical parameters for
which the contrast may be recovered.
\end{abstract}

\pacs{03.75.Dg, 39.20.+q, 03.75.Kk}

\maketitle

\section{Introduction}

The realization of a practical chip-based atom interferometer, using
a Bose-Einstein condensate, (BEC) would revolutionize internal
navigation systems, precision measurements, and, perhaps, quantum
information technology. The production and manipulation of a
condensate using an atom chip is appealing for many reasons. The
trapping potentials, for the same electrical current, have a much
higher frequency then their larger scale counterparts
\cite{reichel02}.  As a result, these devices require less power. An
atomic cloud confined in a trap with a large frequency can be cooled
to ultra-cold temperatures more rapidly \cite{horikoshi05}, making
the realization of a quasi-continuous source of condensed atoms more
feasible.  Chips also enable integration of several relatively
simple components into devices with complex functionality
\cite{hansel01_nature, feenstra04}. Finally, the entire device,
including the vacuum chamber, can be miniaturized \cite{du04},
making chip-based devices the most likely path to practical
applications.

The first experimentally realized BEC-based interferometers used a
double-well potential to manipulate the atoms
\cite{shin04,schumm05}. In this type of interferometer, atoms are
initially prepared in a single potential well that confines the
atoms in all three directions. Then the well is continuously
transformed into a symmetric double well, which must be done
adiabatically slow to avoid collective excitations.  After the
splitting, the wells are physically separated and, due to differences
in the local environment, a quantum phase shift may develop between
the atoms in each well. Nonlinearities, caused by atom-atom
interactions, usually cause problems at the recombination stage in
these types of interferometers.  As a result, the atoms are recombined
by suddenly switching off the trap allowing the two condensates to
ballistically expand, overlap, and interfere. The phase shift can be
measured by fitting the atomic density to a $\cos$ function.

An alternative method for realizing a BEC-based atom interferometer
uses a standing light wave created by two counter-propagating laser
pulses to manipulate the condensate \cite{wang05,garcia06, wu05}.
The BEC is loaded into a waveguide providing confinement along
two directions but not in the third one.  The standing light wave
splits the BEC into two harmonics moving in opposite directions.
A second laser pulse reflects the harmonics by reversing their momentum.
Finally, a third pulse recombines them, thus completing the
interferometer sequence.  Since the atoms are recombined in the
guide, the phase shift between the two arms of the interferometer
can be measured by counting the atoms in the zero momentum state at
the end of the interferometer cycle.

In the first experimental realization of this type of interferometer
\cite{wang05}, the coherence time was about 10~ms. The loss of
coherence was theoretically explained as being due to distortion of
the phase across each atomic cloud \cite{olshanii05}, which is
caused by both the atom-atom interactions and the residual external
parabolic potential in the axial direction.

A similar interferometer with a coherence time of 44~ms has been
realized in Ref.~\cite{garcia06}. The coherence time of this
interferometer was increased in two ways. First, the confining
waveguide was created using macroscopic conductors, instead of
microscopic conductors in a chip-based device. This larger scale
device had looser confinement thus reducing the effect of the
atom-atom interactions.  Additionally, the waveguide was farther
from the conductors minimizing small perturbations due to defects in
the conductors or instabilities in the current. Second, the
interferometer sequence used two reflections, such that each half of
the condensate went through the same path twice in opposite
directions. This method greatly reduces the phase distortion caused by the
residual axial potential.

A third interferometer was recently realized in \cite{horikoshi06}.
This chip-based device has a coherence time of about 15~ms and uses
a sightly different splitting technique. One of the lasers is
frequency-shifted with respect to the other laser, resulting in a
traveling wave optical potential.  Using a $\pi/2$ pulse, the
condensate is split into two harmonics one of which remains at rest
and the other propagates with the momentum $2 \hbar k_{l}$. A $\pi$
pulse acts as a mirror, and finally a second $\pi/2$ recombines the
condensate. As a result of this different splitting mechanism, the
splitting and recombination does not take place in the same physical
location and the device is an atom Mach-Zehnder interferometer. By
varying the radial confinement of the waveguide, the researchers
were able to change the effective strength of the atom-atom
interactions.  They used this to test the theoretical model of
Ref.~\cite{olshanii05} and demonstrated its validity.

In this paper, we discuss possible ways to increase the coherence
time of an atom interferometer using Bragg diffraction to manipulate
the atoms.  Simple analytical expressions for the dynamics of the
interference sequence are derived in the framework of the
hydrodynamic approximation. Their validity is confirmed by their
comparison with direct numerical solution of the Gross-Pitaevskii
equation. We demonstrate that the coherence time can be increased by
changing the recombination time and present simple analytic
expressions for the optimized recombination time and the contrast.

In the rest of the section, we present a simple physical explanation
for the loss of contrast in a BEC Michelson interferometer and
discuss possible ways of restoring it.

The interferometric cycle of duration $T$ starts by illuminating the
motionless BEC cloud $\psi_{0}$ with a splitting pulse from a pair
of counter- propagating laser beams. This pulse acts like a
diffraction grating splitting the cloud into two harmonics
$\psi_{+}$ and $\psi_{-}$. The atoms diffracted into the $+ 1$ order
absorb a photon from a laser beam with the momentum $\hbar k_{l}$
and re-emit it into the beam with the momentum $-\hbar k_{l}$
acquiring the net momentum $2\hbar k_{l}$. The cloud $\psi_{+}$
starts moving with the velocity $v_{0} = 2\hbar k_{l}/M$, where
$k_{l}$ is the wavenumber of the laser beams and $M$ is the atomic
mass. Similarly, the cloud $\psi_{-}$ starts moving with the
velocity $-v_{0}$. The two harmonics are allowed to propagate for
the time $T/2$ and are illuminated by a reflection optical pulse.
The atoms in the harmonics $\psi_{+}$ change their velocity by
$-2v_{0}$ and those in the harmonics $\psi_{-}$ by $2v_{0}$. The
harmonics propagate back for time $T/2$ and are subject to the
action of the recombination optical pulse. After the recombination,
the atoms in general populate all three harmonics $\psi_{0}$ and
$\psi_{\pm}$. The degree of population depends on the relative phase
between the harmonics $\psi_{\pm}$ acquired during the
interferometric cycle and can be used to deduce this phase. In
particular, the wave function of the zero-momentum harmonics
$\psi_{0}$ after the recombination is equal to
\begin{equation}
    \psi_{0} = \frac{1}{\sqrt{2}}(\psi_{+} + \psi_{-}),
\end{equation}
where $\psi_{\pm}$ are the wave functions of the $\pm 1$ harmonics
immediately before the recombination.

Because of the nonlinearity and/or the external potential, the
harmonics $\psi_{\pm}$ do not travel with the velocities $\pm v_{0}$
during the cycle. First, the cloud ``climbing up'' the external
potential slows down and the one moving ``downhill'' speeds up.
Second, because of the nonlinearity, the speeds of the two clouds
after their separation will be slightly larger than $v_{0}$ if the
nonlinearity is repulsive and slightly less than $v_{0}$ if it is
attractive. For definiteness, we shall discuss the influence of the
repulsive nonlinearity assuming that the external potential is zero.
An ideal operation of the interferometer in this case corresponds to
all the atoms populating zero-momentum harmonics $\psi_{0}$ after
the recombination, i.e., to $N_{0} = N_{tot}$.

Because of the atom-atom interaction, the clouds $\psi_{\pm}$ exert
a repulsive force on each other during the time they overlap. This
force accelerates each cloud so that after the separation pulse the
$\psi_{\pm}$  harmonics propagate with velocities $\pm (v_{0} +
\delta v)$, where $\delta v
> 0$. The reflection pulses impart the momenta $\mp 4\hbar k_{l}$ to
the clouds transforming their roles: $\psi_{\pm} \rightarrow
\psi_{\mp}$. After the reflection, the $\pm 1$ harmonics propagate
with the velocities $\pm (v_{0} - \delta v)$. Harmonics'
deceleration due to mutual repulsion during their overlap decreases
the velocity of each harmonics by an additional $\delta v$ so
immediately before the recombination the harmonics' velocities are
$\pm (v_{0} - 2\delta v)$.

The nonzero value of $\delta v$ results in two consequences. First,
since the clouds' speeds after the reflection pulse are smaller than
before the pulse, the $\pm 1$ harmonics at the nominal recombination
time still do not overlap each other completely. This effect is
typically not very significant. Much more important is the fact that
the returning harmonics have momenta that are not equal to $\pm 2
\hbar k_{l}$ and therefore can not be compensated by the
recombination pulse. As a result, the wave function of the
zero-momentum harmonics $\psi_{0}$ after the recombination can be
written as
\begin{equation}
    \psi_{0} \propto \sqrt{n(x)}\cos (\Delta k x),
\end{equation}
where $\Delta k = 2M \delta v/\hbar$  and $n(x)$ are the density
profiles of the harmonics; their possible incomplete overlap has
been neglected. The population of the zero-momentum harmonics is
obtained by the spatial integration of $|\psi_{0}|^{2}$. For $\Delta
k R \ll 1$, where $R$ is the characteristic size of the clouds, all
the atoms after the recombination are indeed in the zeroth
harmonics, i.e. $N_{0} = N_{tot}$. In the opposite case $\Delta k R
\gg 1$, the $\cos$ function oscillates several times across the
cloud and $N_{0}/N_{tot} = 1/2$ resulting in the loss of contrast.
It is worth noting that the accumulation of corrections to the wave
vectors of the clouds is due to the fact that the reflecting pulses
do not reverse the clouds' velocities but rather add a constant
velocity $\pm 2v_{0}$ to them. This explains the fact that the
coherence may be lost due to the presence of an external potential
even when the nonlinearity is negligible.

\begin{figure} 
\includegraphics[width=8.6cm]{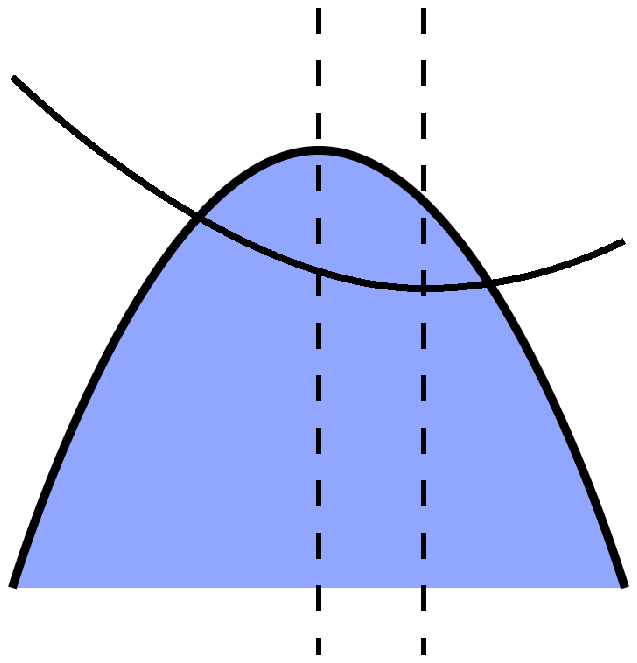}%
\caption{\label{fig:cartoon} The density and phase of the $\psi_{+}$
harmonic before the recombination.}
\end{figure}

The above-discussed loss of coherence due to incomplete cancelation
of the wave vectors of the harmonics by the recombination pulse can
be also visualized in the following way: the wave functions of the
$\psi_{\pm}$ harmonics can be represented as $\psi_{\pm} =
\sqrt{n_{\pm}(x)}\exp(i\phi_{\pm})$, where $\phi_{\pm}$ is the
parabolic phase (the nominal phase $\pm (Mv_{0}/\hbar)x$ is taken
care of by the optical pulses and is not included). In the ideal
situation, the parabolic phase for each cloud is centered at the
middle of the cloud. Nonzero values of $\delta v$ (or, equivalently,
nonzero values of the corrections to the wave vectors of the clouds)
mean that the phase of each cloud leads or lags behind its density
envelope. This situation is schematically illustrated in
Fig.~\ref{fig:cartoon} showing the harmonic $\psi_{+}$ before the
recombination with its phase leading the density envelope. We shall
show that the optimum recombination corresponds to the situation
when the phase profiles, not the density envelopes of the clouds are
on top of each other immediately before the recombination.

Operation with high values of the contrast can be achieved in
several ways. First, the relative magnitudes of the nonlinearity and
the external potential are adjusted in such a way that their effects
cancel each other for a given cycle time $T$ (this is not always
possible). Second, the recombination and/or reflection are conducted
with optical pulses having different wavelength as compared to the
splitting pulse to compensate for the change in the wave vectors of
the moving clouds. Finally, the recombination is carried out not at
the nominal recombination time $T$ but at a time such that $\Delta k
R = 0$. The paper is devoted to the analysis of the last
possibility.

The rest of the paper is structured as follows:
Sec.~{\ref{sec:formulation}} provides general formulation of the
problem, Sec.~\ref{sec:model} introduces analytical model of the
interferometric cycle and Sec.~\ref{sec:signal} is devoted to the
analysis of the contrast and contains the analytical expressions for
the optimized recombination time. These expressions are discussed in
different limiting cases in Sec.~\ref{sec:discussion}.

\section{Formulation of the problem}\label{sec:formulation}
The evolution of the condensate in the interferometer in the
mean-field limit is described by the Gross-Pitaevskii equation
\begin{equation} \label{GPE_initial}
i \hbar \frac{\partial }{\partial t} \Psi({\bm r}, t) = \left( -
\frac{\hbar^2}{2 M} \nabla^2 + V_{tot}({\bm  r}, t) + U_{0} N
|\Psi|^2
 \right) \Psi({\bm  r}, t),
\end{equation}
where $\Psi({\bm  r},t)$ is the wave function of the condensate that
is normalized to one, $N$ is the total number of atoms, $U_0 = 4 \pi
\hbar^2 a_{s}/ M$ characterizes the strength of interatomic
interactions, $a_s$ is the s-wave scattering length and $M$ is the
atomic mass.  The potential $V_{tot} = V_{3D}({\bm  r}, t) +
V_{opt}(t) \cos(2 k_{l} x)$ is the sum of a confining potential
$V_{3D}$ and an optical potential that is created by two
counter-propagating laser beams of wavelength $\lambda = 2 \pi /
k_{l}$, which are detuned from the atomic resonance to avoid
spontaneous emission. The optical potential is used to split,
recombine and reverse direction of propagation of the BEC clouds.

The confining potential is of the form
\begin{equation}\label{V3D}
    V_{3D}({\bm r}, t) = V(x,t) + M \omega_\perp^2 r_\perp^2 / 2,
\end{equation}
where $V(x,t)$ is slowly-spatially-varying potential due to the
environment and $M \omega_\perp^2 r_\perp^2 / 2$ is the guiding
potential providing confinement of the condensate along the two
spatial dimensions ${\bm  r}_\perp = (y,z)$.  In the following we
shall assume that the condensate is tightly confined in the two
transverse dimensions and is in the lowest transverse mode of the
guide
\begin{equation}
\psi_\perp(r_\perp) = \frac{1}{\sqrt{\pi} a_\perp} \exp(-r_\perp^2/2
a_\perp^2),
\end{equation}
where $a_\perp = (\hbar / M \omega_\perp)^{1/2}$ is the transverse
oscillator length.  Factorizing the wave function of the condensate
as $\Psi({\bm r}, t) = \psi(x,t) \psi_\perp( r_\perp)$,
 Eq.~(\ref{GPE_initial}) can be reduced to the one-dimensional equation for the function $\psi(x,t)$.
Introducing dimensionless coordinate $x \rightarrow 2 k_{l} x$ and
time $\tau = t/t_0$, where $t_0 = M/(4 \hbar k_{l}^2)$, this
one-dimensional Gross-Pitaevskii equation can be written as
\begin{equation} \label{GPE_final}
i \frac{\partial}{\partial \tau}\psi(x,\tau) = \left[ -\frac{1}{2}
\frac{\partial^2}{\partial x^2} + v(x,\tau) + \Omega(\tau) \cos(x) +
p |\psi|^2
 \right]\psi(x,\tau),
\end{equation}
where $v = (V/ \hbar)t_{0}$, $\Omega = (V_{opt}/ \hbar)t_{0}$ and $p
= a_{s} N /a_{\perp}^{2} k_{l}$.

The optical potential $\Omega(\tau) \cos x$ acts as a diffraction
grating for the condensate wave function $\psi$.  This grating
diffracts the condensate into several harmonics separated by
multiples of the grating wavevector. If the width of the Fourier
spectrum of the condensate is much smaller than the length of the
grating wavevector (one in our dimensional units), the wave function
$\psi(x,t)$ in Fourier space consists of a series of narrow peaks.
It is therefore convenient to represent $\psi(x,t)$ as
\begin{equation}\label{psi_n_harmonics}
\psi(x,t) = \sum_n \psi_n(x,t) \exp(i n x),
\end{equation}
where harmonics' envelopes $\psi_n(x)$ are slowly-varying functions
of coordinate as compared with the exponentials. The dynamics of
these harmonics are governed by the set of coupled equations
\begin{eqnarray}\label{eqns_for_harmonics}
&&i \left( \frac{\partial}{\partial \tau} + i n
\frac{\partial}{\partial x} \right) \psi_n = \frac{1}{2} \left(-
\frac{\partial^2}{\partial x^2}  + n^2 \right) \psi_n
+ v(x,\tau) \psi_n \nonumber \\
&& + \frac{\Omega(\tau)}{2} (\psi_{n+1} + \psi_{n-1}) +p\sum_{l,m}
\psi_l^* \psi_m \psi_{n-m+l}. \label{eqn:GPE_Modal_form}
\end{eqnarray}

The optical potential $\Omega(\tau)$ in Eq.~(\ref{GPE_final}) is
used to split the initial zero-momentum BEC cloud at the beginning
of the interferometric cycle into the two harmonics with the momenta
$\pm 1$, reverse their direction of propagation in the middle of the
cycle and recombine them at the end. Dynamics of the BEC due to the
optical potential was fully taken into account in solving the
Gross-Pitaevskii equation numerically. In the analytical model,
their action was described in terms of simple transformation
matrices. For self-consistency of the presentation, a brief
derivation of parameters of optical pulses used in the numerical
solution of the Gross-Pitaevskii equation is given in the Appendix.
The results presented in the Appendix have been previously derived
in Refs.~\cite{wu05, wu05_2,garcia06}. The next section is devoted
to the development of analytical model describing evolution of the
BEC between the optical pulses.

\section{Parabolic model}\label{sec:model}
Between the optical pulses the condensate consists of two harmonics
with $n = \pm 1$ whose evolution is described by the set of coupled
equations
\begin{equation}\label{eqns_4_two_harmonics}
i\left(\frac{\partial}{\partial \tau} \pm \frac{\partial}{\partial
x}\right)\psi_{\pm} = -\frac{1}{2}\frac{\partial^{2}}{\partial
x^{2}}\psi_{\pm} + v(x)\psi_{\pm} + p\left(|\psi_{\pm}|^{2} +
2|\psi_{\mp}|^{2}\right)\psi_{\pm}.
\end{equation}
Introducing the density and phase of each harmonic by the relations
\[\psi_{\pm} = \sqrt{n_{\pm}}\exp(i\phi_{\pm})\]
and using the Thomas-Fermi approximation (neglecting the second
derivatives of the density) transforms the set of equations
(\ref{eqns_4_two_harmonics}) to the form
\begin{eqnarray}\label{TF_limit}
    &&\left(\frac{\partial}{\partial \tau} \pm \frac{\partial}{\partial
    x}\right)n_{\pm} = - \frac{\partial}{\partial x}\left(
    n_{\pm}\frac{\partial \phi_{\pm}}{\partial x}\right),\nonumber \\
    &&\left(\frac{\partial}{\partial \tau} \pm \frac{\partial}{\partial
    x}\right)\phi_{\pm} = - \frac{1}{2}\left(\frac{\partial \phi_{\pm}}{\partial x}\right)^{2} -
    v - p(n_{\pm} + 2n_{\mp}).
\end{eqnarray}
We will describe the external potential $v$ by the first two terms
of the Taylor expansion
\begin{equation}\label{potential}
    v(x) = \alpha x + \frac{1}{2}\beta x^{2}
\end{equation}
and analyze the set of Eq.~(\ref{TF_limit}) in the framework of a
parabolic approximation where expressions for both the density and
the pase do not contain terms higher than the second order in
coordinate:
\begin{eqnarray}  \label{TF_density_and_phase}
    n_{\pm} &=& \frac{3}{8R}\left[1 -
    \frac{(x-x_{\pm})^{2}}{R^{2}}\right], \nonumber \\
    \phi_{\pm} &=& \varphi_{\pm} + \kappa_{\pm}(x-x_{\pm}) +
    \frac{g}{2}(x-x_{\pm})^{2}.
\end{eqnarray}
$R$, $x_{\pm}$, $\kappa_{\pm}$, $\varphi_{\pm}$ and $g$ are
functions of time $\tau$ only. The coefficient $3/8$ in the
expression for $n_{\pm}$ follows from the normalization condition
(each harmonics is normalized to $1/2$). Note that for each cloud
its density and phase in Eq.~(\ref{TF_density_and_phase}) are
defined only in the region where the density is nonnegative.
Functions $x_{\pm}(\tau)$ are positions of the centers of mass of
the two moving clouds, $\kappa_{\pm}$ are corrections to their
nominal wavevectors ($\pm 1$) that are due to the external potential
and the nonlinearity and $\varphi_{\pm}$ are the accumulated
coordinate-independent phases. Finally, $R$ is the half-size of each
of the clouds and  the parameter $g$ multiplying the quadratic part
of the phase is analogous to the inverse of the radius of curvature
of the wavefront of a propagating light beam in optics.

Using Eq.~(\ref{TF_density_and_phase}) and the first of
Eq.~(\ref{TF_limit}), one gets
\begin{eqnarray}\label{from_the_first_eqn}
    &&R^{\prime} = g_{\pm}R, \nonumber \\
    &&x_{\pm}^{\prime} = \pm 1 + \kappa_{\pm},
\end{eqnarray}
where the prime means differentiation with respect to time.
Treatment of the second Eq.~(\ref{TF_density_and_phase}) is slightly
complicated by the fact that the regions of existence of $n_{+}$ and
$n_{-}$ do not coincide. Since the functional forms of $n$ and
$\phi$ are fixed, the density profile $n_{\mp}$ should be projected
onto $n_{\pm}$. To do this, one can choose a set of suitable basis
functions defined at the interval $|\xi_{\pm}| \le 1$, where
$\xi_{\pm} = (x - x_{\pm})/R$, that can be used to represent the
density $n_{\pm}$ and the phase $\phi_{\pm}$. The density $n_{\mp}$
should then be expressed in terms of the same basis set retaining
only the functions that describe $n_{\pm}$ and $\phi_{\pm}$. Using
Legendre polynomials $P_{n}(\xi_{\pm})$ as the basis yields
\begin{equation}\label{temp4}
    \frac{16 R}{3}n_{\mp}(\xi_{\pm}) \rightarrow d_{0} \mp d_{1}\xi_{\pm} -
    d_{2}\xi_{\pm}^{2},
\end{equation}
where
\begin{eqnarray}\label{ds}
    &&d_{0} = \left(2 - \frac{7}{2}|q|^{2} + 2|q|^{3} - \frac{1}{8}|q|^{5}\right)\theta(|q| < 2), \nonumber \\
    &&d_{1} = q\left(4 - 3|q| + \frac{1}{4}|q|^{3}\right)\theta(|q| < 2),
    \nonumber \\
    &&d_{2} = \left(2 - \frac{15}{2}|q|^{2} + 5|q|^{3} - \frac{3}{8}|q|^{5}\right)\theta(|q| < 2)
\end{eqnarray}
and $q = (x_{+} - x_{-})/R$. The $\theta$-function in Eq.~(\ref{ds})
is equal to one if its argument is a logical true and zero if its is
a logical false.

Using Eq.~(\ref{temp4}) in the second of Eq.~(\ref{TF_limit}) yields
equations of motion for $g$, $\kappa_{\pm}$ and $\varphi_{\pm}$.
Combining these with Eq.~(\ref{from_the_first_eqn}), we get the
final set of equations
\begin{eqnarray}\label{final_set_of_parab_eqns}
    &&R^{\prime} = g R, \nonumber \\
    &&g^{\prime} = - g^{2} - \beta +
    \frac{3 p }{4 R^{3}}(1 + d_{2}), \nonumber \\
    &&\kappa_{\pm}^{\prime} = - \alpha - \beta x_{\pm} \pm \frac{3
    p}{8 R^{2}}d_{1}, \nonumber \\
    &&x_{\pm}^{\prime} = \pm 1 + \kappa_{\pm}, \nonumber \\
    &&\varphi_{\pm}^{\prime} = \frac{\kappa_{\pm}^{2}}{2} - \alpha x_{\pm}
    - \frac{1}{2}\beta x_{\pm}^{2} - \frac{3 p}{8R}(1 + d_{0}),
\end{eqnarray}
where prime means differentiation with respect to time. Equations
(\ref{final_set_of_parab_eqns}) have simple physical interpretation.
The rates of change of the coordinates of the two clouds
$\psi_{\pm}$ are given by the relations $x_{\pm}^{\prime} = \pm 1 +
\kappa_{\pm}$, i.e., the clouds move with velocities $\pm 1 +
\kappa_{\pm}$. The major contributions to the velocities $\pm 1$ are
due to the momenta imparted to the clouds by the optical pulses. The
corrections $\kappa_{\pm}$ are due to the external potential
(parameters $\alpha$ and $\beta$) and the nonlinearity. The cloud
``climbing up'' the external potential slows down and the one moving
``downhill'' speeds up. If the nonlinearity is repulsive ($p > 0$),
the speeds of the two clouds after their separation will be slightly
larger than one and if it is attractive, slightly less than one. The
functions $d_{0}$, $d_{1}$ and $d_{2}$ given by Eq.~(\ref{ds})
describe mutual interaction of the two clouds. They depend on the
relative displacement of the clouds $q = (x_{+} - x_{-})/R$ and are
nonzero only when $|q| < 2$, i.e., when the clouds overlap. The
other terms containing the nonlinearity parameter $p$ describe self
interaction for each of the clouds and are always nonzero.

\subsection{Evolution of $\kappa_{\pm}$ and
$x_{\pm}$}\label{subsec:x_and_v}
During the interferometric cycle the two BEC clouds $\psi_{\pm}$ may
be partially overlapping or non-overlapping. In the subsequent
analysis, it will be assumed that the size of each cloud does not
change significantly at the time intervals $\tau \propto R$ that it
takes for the clouds to pass each other. The conditions of
applicability of this assumption are given by
Eq.~(\ref{applicability_R_does_not_change}). Additionally, it will
be assumed that $\beta T^{2} \ll 1$, where $T$ is the duration of
the interferometric cycle.

Time evolution of $\kappa_{\pm}$ and $x_{\pm}$ is governed by the
set of two coupled equations (cf. (\ref{final_set_of_parab_eqns}))
\begin{eqnarray}\label{v_and_x}
    &&\kappa_{\pm}^{\prime} = - \alpha - \beta x_{\pm} \pm \frac{3p}{8 R^{2}}d_{1}, \nonumber \\
    &&x_{\pm}^{\prime} = \pm 1 + \kappa_{\pm},
\end{eqnarray}
 Solution of Eqs.~(\ref{v_and_x}) can be written as
\begin{eqnarray}\label{v_pm_simple3}
    \kappa_{\pm}(\tau) = \kappa_{\pm,0} -\alpha \tau -x_{\pm, 0}\beta \tau \mp
    \frac{1}{2}\beta \tau^{2}
    \pm \frac{3p}{16R}\int_{q_{0}}^{q_{0} + 2 \tau/R_{0}}dq d_{1}(q),
\end{eqnarray}
\begin{eqnarray}\label{x_pm_simple3}
    &&x_{\pm}(\tau) = x_{\pm, 0}  + (\pm 1 + \kappa_{\pm,0}) \tau -
    \frac{1}{2}\alpha \tau^{2} - \frac{1}{2}x_{\pm,0}\beta \tau^{2} \mp \frac{1}{6}\beta \tau^{3} \nonumber \\
    &&\pm \frac{3p}{32}\int_{q_{0}}^{q_{0} + 2 \tau/R_{0}} dq \int_{q_{0}}^{q}dq^{\prime}d_{1}(q^{\prime}),
\end{eqnarray}
where $\kappa_{\pm,0}$, $x_{\pm,0}$ and $R_{0}$ are initial values
of $\kappa_{\pm}$, $x_{\pm}$ and $R$. In deriving
Eq.~(\ref{v_pm_simple3}) and (\ref{x_pm_simple3}), the dynamics of
the relative separation between the clouds in evaluating function
$d_{1}$ was approximated by the relation
\begin{equation}\label{q_of_tau}
    q(\tau) = q_{0} + \frac{2\tau}{R_{0}},
\end{equation}
i.e., the terms with $\kappa_{\pm}$ were neglected as compared to
one in evaluating $q(\tau)$.

The interferometric cycle of duration $T$ starts by applying the
splitting optical pulses to the motionless cloud $\psi_{0}$, letting
harmonics $\psi_{\pm}$ propagate for the time $T/2$, reverse their
directions of propagation by applying the reflection pulses, letting
the harmonics $\psi_{\pm}$ evolve for the time $T/2$ and apply the
recombination optical pulses.

Immediately after the splitting pulses at $\tau = 0$, the center of
mass of each harmonic is $x_{\pm,0} = {0}$, $q_{0} = 0$ and
$\kappa_{\pm,0} = 0$. The reflection pulse reverses directions of
propagation of the two harmonics by adding momenta $\mp 2$ to the
momenta $\pm 1 + \kappa_{\pm}$ of $\psi_{\pm}$. After the reflection
pulse the harmonic $\psi_{+}$ becomes $\psi_{-}$ and vice versa. As
a result, immediately after the reflection pulse, $x_{\pm, 0} =
x_{\mp}(T/2)$ and $\kappa_{\pm, 0} = \kappa_{\mp}(T/2)$.

At the nominal recombination time $\tau = T$, the corrections to the
velocities $\kappa_{\pm}$ and the center of mass coordinates
$x_{\pm}$ are given by the relations
\begin{eqnarray} \label{v_Final}
    \kappa_{\pm}(T) &=& - \alpha T \pm \frac{1}{4} \beta T^2 \mp p
    \left[ \frac{1}{R_0} D_{1}(T/R_{0}) + \frac{1}{R_T} D_1(T/R_T)\right],
\end{eqnarray}

\begin{eqnarray} \label{x_Final}
    &&x_{\pm}(T) = -\frac{1}{2} \alpha T^2 \pm \frac{1}{8} \beta T^3 \nonumber \\
    && \mp \frac{p}{2} \left[ \int_{T/R_T}^{T/R_0} dq D_1(q) + \frac{T}{R_0}
    D_1(T/R_0) + \frac{T}{R_T} D_1(T/R_T) \right],
\end{eqnarray}
where
\begin{equation}\label{D1}
    D_1(x) =
    \left\{\begin{array}{ccc}
    \frac{3}{16} x^2 \left(2 - x + \frac{1}{20} x^3 \right) &,& x < 2 \\
    D_{1}(2) = 3/10 &, & x > 2,
\end{array}
\right.
\end{equation}
$R_0$ is the size of the harmonics at the separation stage and $R_T$
is the size during recombination.
\subsection{Evolution of $g$ and $R$}
Evolution of $g$ and $R$ is governed by the set of two coupled
equations (see Eq.~(\ref{final_set_of_parab_eqns}))

\begin{eqnarray}\label{g_and_R}
    &&R^{\prime} = g R, \nonumber \\
    &&g^{\prime} = - g^{2} - \beta +
    \frac{3 p }{4 R^{3}}(1 + d_{2}).
\end{eqnarray}

The explicit expressions for $g$ and $R$ at time intervals $\tau$
such that $R$ does not change significantly, i.e., $|\Delta R| \ll
R$, are of the form

\begin{eqnarray}\label{g_when_R_does_not_change}
    g(\tau) = g_{0} - \beta \tau + \frac{3p}{8R_{0}^{2}}\int_{q_{0}}^{q_{0} + 2 \tau/R_{0}}d q[1 +
    d_{2}(q)],
\end{eqnarray}
\begin{equation}\label{R_when_R_does_not_change}
        R(\tau) = R_{0} + R_{0}\left[g_{0} \tau - \frac{\beta}{2}\tau^{2} +
        \frac{3p}{8R_{0}^{2}}\int_{0}^{\tau}d\tau^{\prime}\int_{q_{0}}^{q_{0} + 2\tau^{\prime}/R_{0}}dq[1 +
        d_{2}(q)]\right],
\end{equation}

where $g_{0} = g(0)$, $q_{0} = q(0)$ and $R_{0} = R(0)$ are initial
values of $g$, $q$ and $R$. In deriving
Eqs.~(\ref{g_when_R_does_not_change}) and
(\ref{R_when_R_does_not_change}), the terms of the order
$\kappa_{\pm}$ in the equations for $x_{\pm}$ have been neglected as
compared to one. The dynamics of the relative separation between the
clouds in the framework of this approximation is given by the
expression
\begin{equation}
    q(\tau) = q_{0} + \frac{2\tau}{R_{0}}
\end{equation}
Equations (\ref{g_when_R_does_not_change}) and
(\ref{R_when_R_does_not_change}) are valid provided
\begin{equation}
   g_{0}\tau, \; \beta \tau^{2},\; \frac{p\tau^{2}}{R_{0}^{3}} \ll 1
\end{equation}
In the analysis of Sec.\ref{subsec:x_and_v} and in the rest of the
paper it is assumed that the size of each cloud does not change
significantly during the time $\tau = R$ that it takes for the
clouds to pass each other. The conditions of applicability of this
approximation are
\begin{equation}\label{applicability_R_does_not_change}
    g_{0}R_{0}, \; \beta R_{0}^{2},\;
    \frac{p}{R_{0}} \ll 1
\end{equation}
Using Eq.~(\ref{g_when_R_does_not_change}), we get the following
expression for the value of $g$ at the recombination time in the
limit when $R$ does not change significantly during the
interferometric cycle:

\begin{eqnarray}\label{g_T_when_R_does_not_change}
    g(T) = g_{0} - \beta T + \frac{3pT}{4R^{3}} +
    \frac{p}{R^{2}}D_{2}(T/R),
\end{eqnarray}
where
\begin{equation}\label{D2}
    D_{2}(x) = \left\{\begin{array}{ccc}
    \frac{3}{4}x\left(2 - \frac{5}{2}x^{2} + \frac{5}{4}x^{3} -
    \frac{1}{16}x^{5}\right) &,& x < 2, \\
    D_{2}(2) = 0 &,& x > 2
    \end{array}\right.
\end{equation}

The limit $|\Delta R| \ll R$ can correspond to both $\tau < R$ when
the clouds stay overlapped during all the cycle and to $\tau \gg R$
when they do not overlap most of the cycle. The second limit of
interest to be considered in this section $\tau \gg R$ explicitly
deals with the situation when the clouds do not overlap most of the
time. In this limit, the contribution coming from the function
$d_{2}$ in Eq.~(\ref{g_and_R}) (interaction between the clouds) can
be neglected as compared to their self action. The function $g$ in
this limit is given by the relation
\begin{equation}\label{g_longT}
    g(\tau) = \frac{1}{r} \frac{d}{d\tau} r =  \frac{\mbox{sign}~g_0}{r} \left[
    g_0^2 - \frac{3 p}{2 R_0^3} \left(\frac{1}{r} - 1 \right) - \beta (r^2 -1)
    \right]^{1/2},
\end{equation}
where $r = R(\tau)/R_0$. Note that Eq.~(\ref{g_longT}) is valid for
any values of $R(\tau)$. The general expression for $R(\tau)$ can be
obtained in terms of elliptic integrals but is too cumbersome to be
of practical use. In the limit where the relative change in the size
of each harmonic is small($|r - 1| \ll 1$), one gets
\begin{eqnarray}\label{R_g_approx}
    &&R(\tau) = R_{0} + R_{0}\left[g_{0} \tau + \left(\frac{3p}{4R^{3}_{0}} -
    \beta\right)\frac{\tau^{2}}{2}\right], \nonumber \\
    &&g(\tau) = g_{0} + \left(\frac{3p}{4R^{3}_{0}} -
    \beta\right)\tau.
\end{eqnarray}
These expressions coincide with
Eqs.~(\ref{g_when_R_does_not_change}) and
(\ref{R_when_R_does_not_change}) when $\tau \gg R$ and describe the
situation when the clouds do not overlap most of the time but their
sizes do not changes significantly during all their evolution time.

In the opposite limit $R(\tau) \gg R_{0}$,
\begin{equation}
    g(\tau) = {\rm sign}~g_{0}\frac{R_{0}}{R(\tau)}\left[g_{0}^{2} + \frac{3p}{2R_{0}^{3}} - \beta \frac{R^{2}(\tau)}{R_{0}^{2}}\right]^{1/2}
\end{equation}
\subsection{Evolution of $\varphi_{\pm}$}

In an interferometric experiment, the quantity of interest is not
the absolute phase of each harmonic $\varphi_{\pm}$, but rather the
relative phase $\Delta \varphi = \varphi_{+} - \varphi_{-}$. The
time evolution of $\Delta \varphi$ is governed by the equation
\begin{equation}
    \Delta \varphi^{\prime} = \frac{1}{2} (\kappa_{+}^2 - \kappa_{-}^2 ) - \alpha (x_+ - x_-)
    - \frac{1}{2} \beta (x_+^2 - x_-^2)
\end{equation}
Using results of Section \ref{subsec:x_and_v} and neglecting terms
containing products and quadratic or higher combinations of
$\alpha$, $\beta$ and $p$ yields
\begin{equation}\label{Delta_varphi_recombination}
    \Delta \varphi(T) = - \frac{\alpha}{2} T^{2}
\end{equation}

\section{The interference signal}\label{sec:signal}
%
The wavefunction of the zero-momentum harmonics after the
recombination is given by the expression:
\begin{eqnarray}\label{psi_after_recombination_general}
    &&\psi_{0}(x) = \frac{1}{\sqrt{2}}\left[\psi_{+}(x) +
    \psi_{-}(x)\right]
    \propto \frac{1}{\sqrt{2}}\left[\sqrt{n(x-x_{+})}\exp(i \theta_{0}/2 + i\Delta k x/2) \right. \nonumber \\
    &&+ \left. \sqrt{n(x+x_{+})}\exp(-i\theta_{0}/2 -i\Delta k x/2)\right].
\end{eqnarray}
Here $\psi_{\pm}$ are the wavefunctions of the $\pm 1$ harmonics
before the recombination,
\begin{equation}\label{n_after_recombination}
    n(x) = \frac{3}{8R}\left(1 - \frac{x^{2}}{R^{2}}\right),
 \end{equation}
\begin{equation}\label{phase_after_recombination}
    \theta_{0} = (\varphi_{+} - \varphi_{-}) -
    (\kappa_{+}x_{+} - \kappa_{-}x_{-}) + \frac{g}{2}(x_{+}^{2} -
    x_{-}^{2})
\end{equation}
and
\begin{equation}\label{deltak_recombination}
    \Delta k  = \Delta \kappa - g\Delta x,
\end{equation}
where $\Delta \kappa = \kappa_{+} - \kappa_{-}$ and $\Delta x  =
x_{+} - x_{-}$. All quantities in
Eq.~(\ref{phase_after_recombination}) are evaluated at the
recombination time.

If the density envelopes of the $\pm 1$ harmonics sufficiently
overlap at the recombination stage,
Eq.~(\ref{psi_after_recombination_general}) can be simplified to

\begin{equation}\label{psi_after_recombination}
    \psi_{0}(x) = \sqrt{n(x)}\cos(\theta_{0}/2 + \Delta k x/2)
\end{equation}
Population of the zero-momentum harmonics $N_{0}$ is given by the
spatial integration of $|\psi_{0}|^{2}$ yielding
\begin{equation}\label{population_of_zeroth_harmonics}
    \frac{N_{0}}{N_{tot}} = \frac{1}{2}\left(1 + V\cos \theta_{0}\right),
\end{equation}
where the contrast of the interference fringes $V$ is given by the
expression
\begin{equation}\label{contrast_simple}
    V = \frac{3}{(\Delta k R)^{3}}[\sin (\Delta k R) - \Delta k
    R\cos(\Delta k R)].
\end{equation}
For $\Delta k R \ll 1$, the population of the zero-momentum state is
given by the relation
\[
    N_{0}/N_{tot} = \cos^{2}(\theta_{0}/2).
\]
In this limiting case the population depends on the relative
accumulated coordinate-independent phase $\theta_{0}$ between the
two $BEC$ clouds and exhibits interference fringes as a function of
this phase.

 In the opposite case $\Delta k R \gg 1$, the $\cos$ function
 in Eq.~(\ref{psi_after_recombination}) oscillates several times
 across the cloud and
\[
    N_{0}/N_{tot} = 1/2
\]
independently of the value of the relative phase shift.

\begin{figure} 
\includegraphics[width=8.6cm]{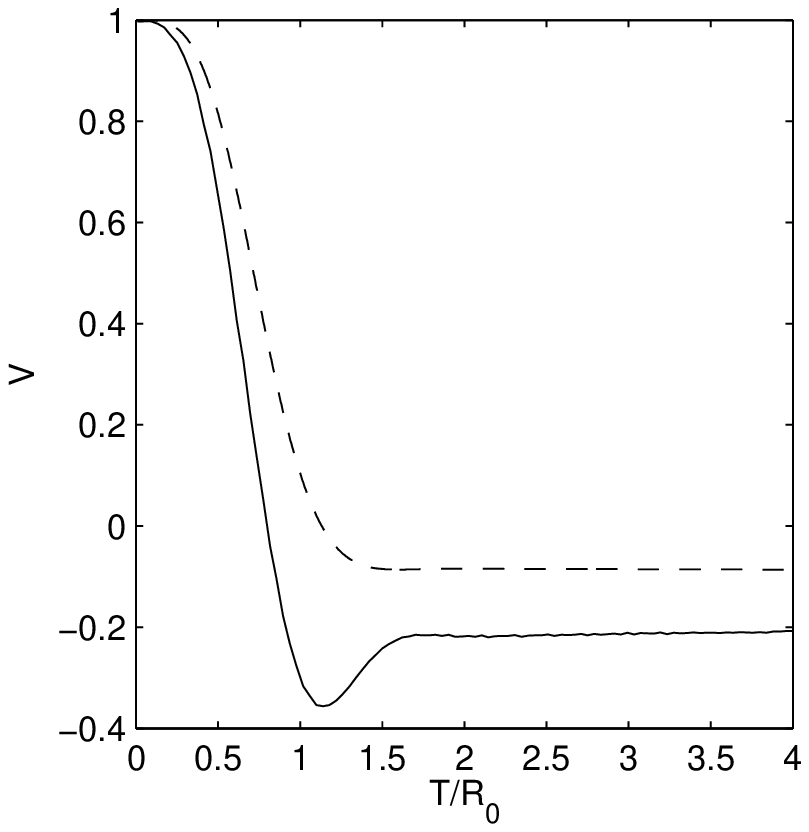}%
\caption{\label{fig:N0_vs_T} The contrast $V$ versus the cycle time
$T/R_{0}$ for $R_0=500$, $p=5$, $\alpha = 0$, $g_{0} = 0$ and $\beta
= 0$.}
\end{figure}

Equations (\ref{x_Final}) and (\ref{v_Final}) show that both the
nonlinearity of the condensate $p$ and the quadratic contribution to
the external potential $\beta$ can result in nonzero values of
$\Delta k$ given by Eq.~(\ref{deltak_recombination}) and thus be
responsible for the loss of interferometric contrast as illustrated
by Fig.~\ref{fig:N0_vs_T}. This figure shows the contrast $V$
defined by the relation  $N_{0}/N_{tot} = (1 + V)/2$, where
$N_{0}/N_{tot}$ is the relative population of the zero-momentum
harmonics at the end of the interferometric cycle, as a function of
the ratio of the cycle time to the initial size of the harmonic
$T/R_0$.  The solid line corresponds to the numerical solution of
the Gross-Pitaevskii equation (\ref{GPE_final}). The dashed line is
given by Eq.~(\ref{contrast_simple}), where $\Delta k R$ is
calculated with the help of analytical expressions
(\ref{deltak_recombination}), (\ref{v_Final}), (\ref{x_Final}) and
(\ref{g_T_when_R_does_not_change}).

\begin{figure} 
\includegraphics[width=8.6cm]{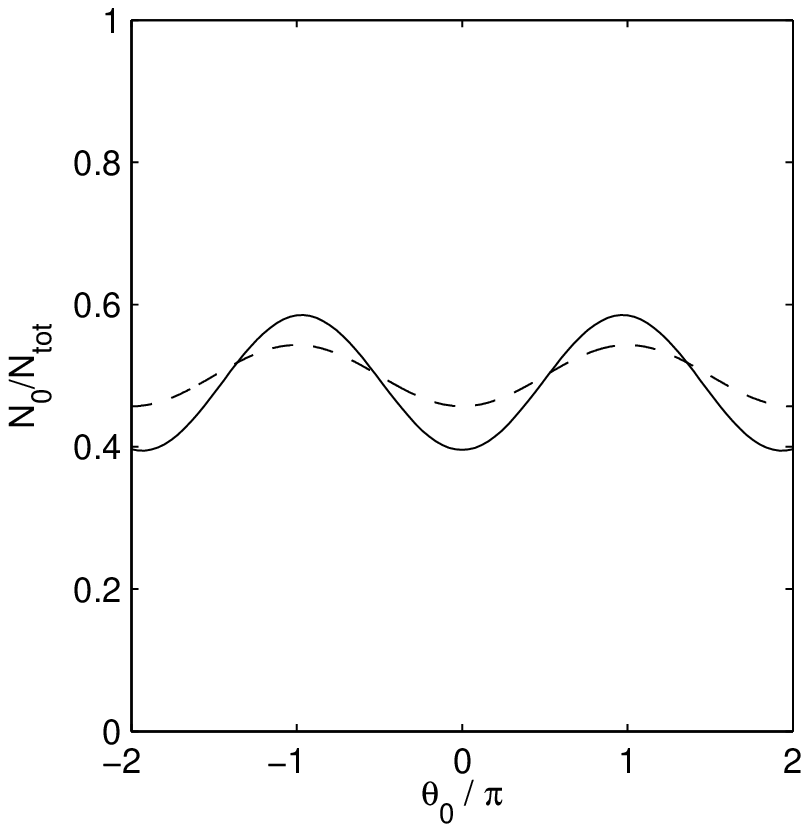}%
\caption{\label{fig:fringes_nominal} Relative population of the
zeroth-momentum harmonic $N_{0}/N_{tot}$ versus the relative
accumulated phase $\theta_{0} = -\alpha T^{2}/2$ for $T = 2000$.
Other parameters (except nonzero values of $\alpha$) are as in
Fig.\ref{fig:N0_vs_T}.}
\end{figure}

Since the linear slope of the potential is zero ($\alpha = 0$),
$\theta_{0} = 0$. Equation (\ref{contrast_simple}) then predicts
that for $\Delta k R = 0$, $V = 1$. As is seen in
Fig.~\ref{fig:N0_vs_T}, the contrast indeed equals one for short
cycles (small $T$). Larger values of $T$ correspond to larger
interaction times between the two clouds and an increase in $\Delta
k R$ due to this interaction. As the interaction time increases, the
contract $V$ given by Eq.~(\ref{contrast_simple}) goes down from one
to small negative value resulting in the values of $N_{0}/N_{tot}$
slightly below 1/2. At times larger than about $T/R_{0} = 1.5$ the
two harmonics completely pass each other and stop overlapping during
a part of the cycle. The interaction time between the harmonics (the
time when they overlap) is now smaller than the cycle time and does
not depend on it. The contrast and the population of the
zero-momentum harmonic reach their limiting values.
%
%
Figure \ref{fig:fringes_nominal} shows the dependence of the
population of the zeroth-order harmonic after the recombination
$N_{0}/N_{tot}$ on the relative accumulated phase shift $\theta_0 =
-\alpha T^2/2$. The solid line corresponds to the numerical solution
of the Gross-Pitaevskii equation (\ref{GPE_final}). The dashed line
is Eq.~(\ref{population_of_zeroth_harmonics}) with $\Delta k R$
given by Eqs.~(\ref{v_Final}), (\ref{x_Final}) and
(\ref{g_T_when_R_does_not_change}). As is seen in
Fig.~\ref{fig:N0_vs_T}, the cycle time $T/R_{0} = 4$ corresponds to
small negative values of the contrast ($V \approx -0.2$ as given by
the Gross-Pitaevskii equation and $V \approx -0.1$ as given by the
analytic model). Low values of the contrast result in the washout of
the interference fringes shown in Fig.~\ref{fig:fringes_nominal}. It
also should be noted that since the contrast $V$ is negative, the
symmetric recombination with $\theta_{0} = 0$ corresponds not to the
maximum, but the minimum population $N_{0}$ of the zero-momentum
harmonic.

\begin{figure} 
\includegraphics[width=8.6cm]{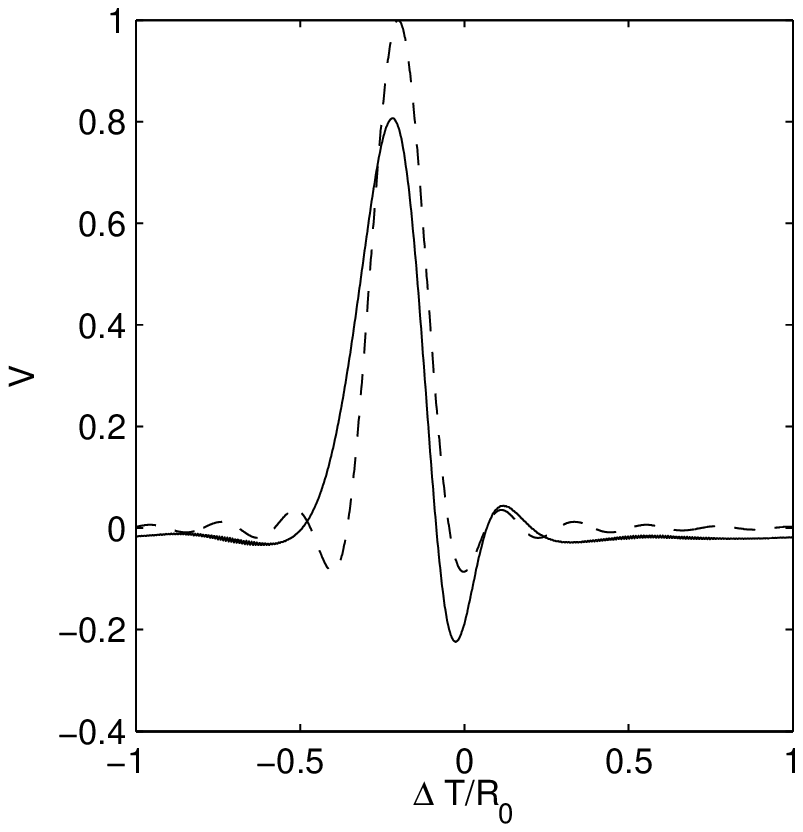}%
\caption{\label{fig:N0_vs_dT} The contrast $V = 2N_{0}/N_{tot} - 1$
as a function of $\Delta T/R_{0}$.}
\end{figure}

\begin{figure} 
\includegraphics[width=8.6cm]{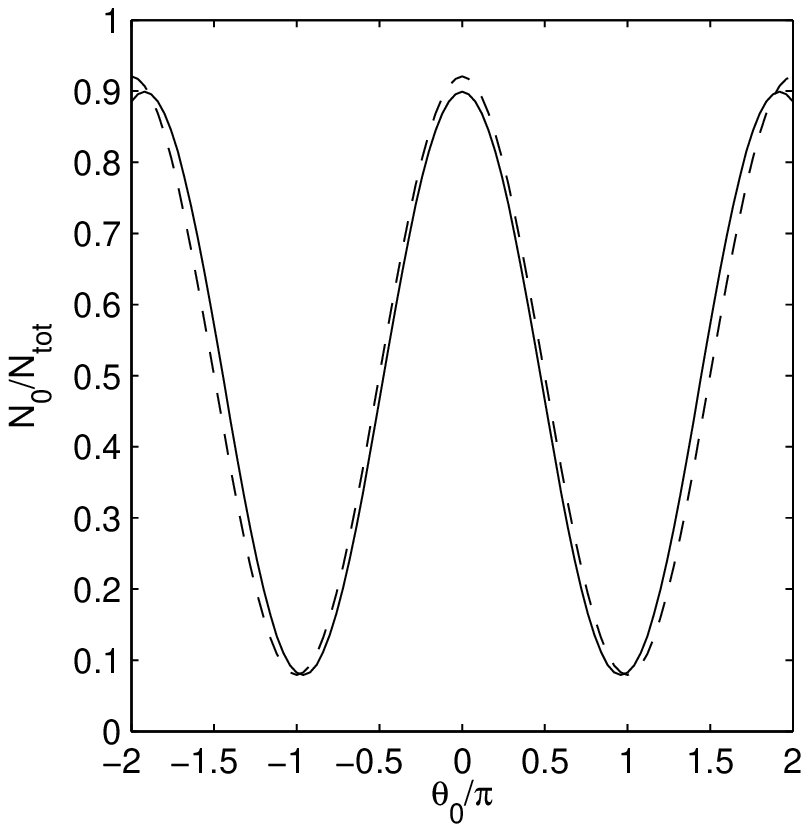}%
\caption{\label{fig:fringes_improved} Relative population of the
zero-order harmonic $N_{0}/N_{tot}$ versus  the relative accumulated
phase shift $\theta_{0}$ for $T/R_{0} = -0.2$. All other parameters
are the same as for Fig.~\ref{fig:fringes_nominal}}.
\end{figure}

Figures \ref{fig:N0_vs_T} and \ref{fig:fringes_nominal} demonstrate
that recombination with nonzero value of the linear wavevector
$\Delta k$ (see Eq.~(\ref{deltak_recombination})) washes out the
interference fringes.  Since $\Delta k$  is a function of time, this
effect can be compensated for by conducting the recombination not at
the nominal time $T$ but at a slightly different time $T + \Delta T$
when $\Delta k R = 0$ (in general, $\Delta T$ may be both positive
and negative). Figure \ref{fig:N0_vs_dT} shows the contrast $V =
2N_{0}/N_{tot} - 1$, where $N_{0}/N_{tot}$ is relative population of
the zero-momentum harmonic, as a function of the time $\Delta T$.
Negative (positive) values of $\Delta T$ correspond to the
recombination taking place slightly before (after) the nominal
recombination time $T$. The parameters for Fig.~\ref{fig:N0_vs_dT}
are $T = 2000$, $R_{0} = 500$ and  $p=5$ with all other parameters
being zero. The solid line is the solution of the Gross-Pitaevskii
equation and the dashed line is obtained with the help of
Eqs.~(\ref{population_of_zeroth_harmonics}), (\ref{v_Final}),
(\ref{x_Final}) and (\ref{g_longT}). Recombination at the nominal
time $\Delta T = 0$ corresponds to a small value of the contrast and
a washout of the fringes as is shown in
Fig.~\ref{fig:fringes_nominal}.
%
%
Figure \ref{fig:N0_vs_dT} indicates that if the recombination takes
place at $\Delta T/R_{0} \approx -0.2$, the contrast of the fringes
becomes much larger. This is confirmed by
Fig.~\ref{fig:fringes_improved}, which shows $N_{0}/N_{tot}$ versus
the relative accumulated phase shift $\theta_{0} = - \alpha [ (T +
\Delta T)^2 -  T^2/2]$  for $\Delta/R = -0.2$ and all other
parameters the same as in Fig.~\ref{fig:fringes_nominal}. The solid
line is the solution of the Gross-Pitaevskii equation and the dashed
line is the result of the numerical solution of
Eqs.~(\ref{final_set_of_parab_eqns}).
%

The simple estimate using the condition $(\Delta k R)(T + \Delta T)
= 0$ yields
\begin{equation}\label{Delta_T}
 \frac{\Delta T}{R_{T}} = -\frac{(\Delta kR)_{T}}{R(\Delta k R)^{\prime}_{T}}.
\end{equation}
The ratio $\Delta T/R$ gives the relative displacement of the two
clouds at the recombination time $T + \Delta T$ since the clouds
pass across each other in time $R$ (each cloud has the size $2R$ and
the relative speed is $2$).

The population of the zero-momentum harmonics depends not only on
the magnitude of $\Delta k R$, but on the degree of overlap of the
two density envelopes at the recombination time (see
(\ref{psi_after_recombination_general})). Estimate (\ref{Delta_T})
takes into account only changes in $\Delta k R$ but not in the
overlap in evaluating $\Delta T$.  The last can be taken into
account in the framework of
Eq.~(\ref{psi_after_recombination_general}) at the expense of making
formulas more cumbersome and turn out to be not very significant. As
we shall see, Eq.~(\ref{Delta_T}) is in a very good qualitative and
quantitative agrement with the results of numerical solution of the
Gross-Pitaevskii equation (\ref{GPE_final}). Finally it should be
mentioned that nonzero values of $\Delta T/R$ mean incomplete
overlap and thus the contrast less than one even at the optimized
time. The larger is $|\Delta T|/R$, the smaller the contrast. The
estimate (\ref{Delta_T}) implicitly implies that $|\Delta T|/R \le
1$ because correction to the recombination time is meaningful only
if the clouds overlap at the time $T + \Delta T$. If, for some set
of parameters, estimate (\ref{Delta_T}) yields $|\Delta T|/R > 1$,
the coherence can not be recovered for this set of parameters.

Using the explicit expressions for $(\Delta k R)$ and its time
derivative obtained with the help of
Eqs.~(\ref{deltak_recombination}) and
(\ref{final_set_of_parab_eqns}) results in the relation
\begin{equation}\label{Delta_T_final}
    \frac{\Delta T}{R} = \frac{\Delta \kappa - g\Delta x}{2gR - 3p\Delta
    x/4R^{2}}.
\end{equation}
In Equation~(\ref{Delta_T_final}), $\Delta \kappa(T)$ and $\Delta
x(T)$ are evaluated using Eq.~(\ref{v_Final}) and (\ref{x_Final}).
The function $g(T)$ should be evaluated using several different
expressions depending on the parameters of the problem. For $|\Delta
R| \ll R$, $g(T)$ is given by
Eq.~(\ref{g_T_when_R_does_not_change}). In this case $R_{0} = R_{T}
= R$. If $T \gg R_{0}$, $g(T)$ is given by Eq.~(\ref{g_longT}).
Since we are assuming that $|\Delta R| \gg R$ at times it takes the
clouds to pass through each other,
Eqs.~(\ref{g_T_when_R_does_not_change}) and (\ref{g_longT}) cover
all possible situations. If $|\Delta R|$ is not small as compared to
$R$, the size of the clouds $R_{T}$ at the end of the cycle should
be evaluated by numerical integration of Eq.~(\ref{g_longT}).

Equation (\ref{Delta_T_final}) is relatively complex because it
covers both the case when the size of the clouds does not change
significantly during the cycle and the opposite limit when the final
size is much larger than the initial one. All the relevant physics
can be understood by discussing the case $|\Delta R| \ll R$ when
Eq.~(\ref{Delta_T_final}) acquires especially simple form
\begin{equation}\label{Delta_T_simple}
    \frac{\Delta T}{R} = \frac{\Delta \kappa}{2gR} = -
    \frac{1}{4}\frac{8D_{1}(\zeta) - (\beta R^{3}/p)\zeta^{2}}{3\zeta/4 + D_{2}(\zeta) +  g_{0}R^{2}/p - (\beta
    R^{3}/p)\zeta},
\end{equation}
where $\zeta = T/R$.

The contrast at the optimized recombination time can be evaluated by
accounting for an incomplete overlap of the clouds using
Eq.~{\ref{psi_after_recombination_general}} and is given by the
approximate expression
\begin{equation}\label{contrast_optimized}
    V \approx 1 - \frac{3}{2}\left(\frac{\Delta T}{R}\right)^{2}\left[\ln
    \frac{R}{|\Delta T|} + 2\ln2 - \frac{1}{2}\right].
\end{equation}
Equations (\ref{Delta_T_simple}) and (\ref{contrast_optimized}) are
the main analytical results of the paper. In
Section~\ref{sec:discussion}, they will be analyzed in several
illustrative cases.

\section{Discussion}\label{sec:discussion}

\subsection{Influence of the nonlinearity $p$ for $g_{0} = \beta = 0$}
%
For $\beta = g_{0} = 0$, Eq.~(\ref{Delta_T_simple}) becomes
\begin{equation}\label{Delta_T_g_and_beta_zero}
    \frac{\Delta T}{R} = \frac{- 2 D_{1}(T/R)}{D_{2}(T/R) + 3T/4R}
\end{equation}
Equation (\ref{Delta_T_g_and_beta_zero}) shows that the correction
to the recombination time depends only on the single parameter $T/R$
and does not depend on the nonlinearity of the condensate $p$ (the
applicability of the parabolic approximation requires $pR \gg 1$).
This is due to the fact that both the $\Delta \kappa$ and the $gR$
terms are proportional to the nonlinearity parameter $p$. At small
values of $T/R$, $D_{1}(T/R) \approx (3/8)(T/R)^{2}$, $D_{2}(T/R)
\approx (3/2)(T/R)$ (cf. Eqs.~(\ref{D1}) and (\ref{D2})) so
\begin{equation}\label{Delta_T_g_and_beta_zero_zeta_small}
 \frac{\Delta T}{R} = -\frac{3}{10}\frac{T}{R},
\end{equation}
i.e., $\Delta T/R$ grows linearly with $T/R$. The correction to the
recombination time $\Delta T/R$ reaches maximum for $T/R \approx 2$
when the duration of the cycle is such that the two clouds at their
maximum separation stop overlapping. At longer cycle times $T/R >
2$, both $D_{1}$ and $D_{2}$ become constants and $\Delta T/R$
starts decreasing inversely proportional to $T$:
\begin{equation}\label{Delta_T_g_and_beta_zero_zeta_large}
    \frac{\Delta T}{R} = - \frac{4}{5}\left(\frac{T}{R}\right)^{-1}.
\end{equation}
This behavior has simple physical explanation. The difference
between the corrections to the propagation velocities of the clouds
$\Delta \kappa$ is due to the nonlinear interaction between the
clouds and is accumulated only when the clouds overlap (see the
definition of $D_{1}$ Eq.~(\ref{D1})). For short cycle times $T/R <
2$, when the clouds overlap during all the cycle, the nonlinear
effects are accumulated during all times and $\Delta \kappa \propto
T^{2}$. The parabolic phase described by the coefficient $g(T)$
grows linearly with time $T$, so the correction to the recombination
time $T$ is a growing function of the cycle time $T$. For $T > R$,
when the clouds fully separate during the cycle, $\Delta \kappa$ is
at its maximum possible value and stops growing further. The
quadratic phase profile of each cloud, on the other hand, keeps
growing as a function of time, i.e., $g$ becomes larger, thus
resulting in the decrease of $\Delta T$.

\begin{figure} 
\includegraphics[width=8.6cm]{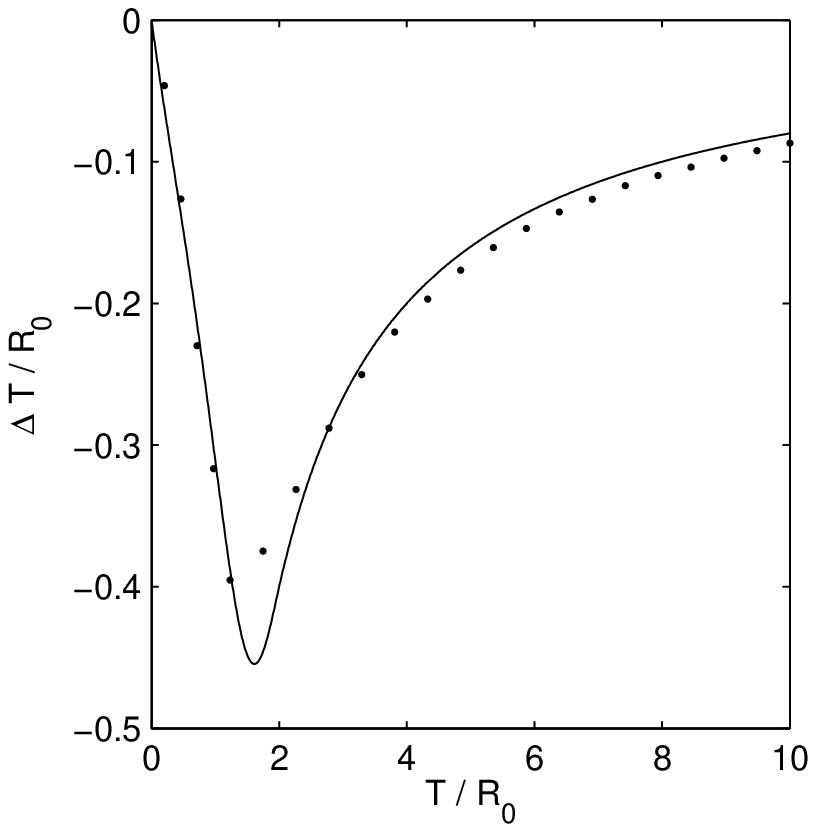}%
\caption{ \label{fig:DeltaT_vs_T} The shift in the recombination
time $\Delta T/R_{0}$ as a function of the nominal recombination
time $T/R_{0}$ for $R_0 = 500$ and $p=5$. The dots are the results
of the numerical solution of the GPE and the solid line is the
analytical model.}
\end{figure}

\begin{figure} 
\includegraphics[width=8.6cm]{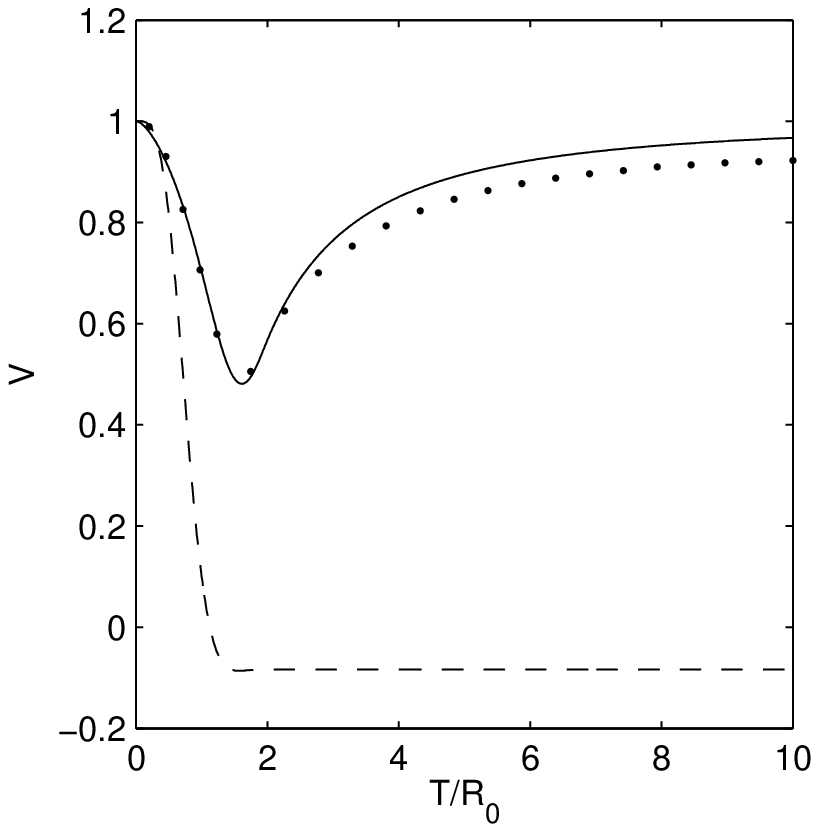}%
\caption{ \label{fig:contrast_vs_T} Optimized contrast of the
interference fringes $V$ for the parameters of
Fig.~\ref{fig:DeltaT_vs_T}. The dots are the results of the
numerical solution of the GPE and the solid line is the analytical
model. The dashed line is the contrast at the nominal recombination
time.}
\end{figure}

The dependence of the shift in the recombination time $\Delta
T/R_{0}$ on the cycle time $T/R_{0}$ is shown in
Fig.~\ref{fig:DeltaT_vs_T} for $R_{0} = 500$ and $p = 5$. The
maximum cycle times shown in Fig.~\ref{fig:DeltaT_vs_T} correspond
to the maximum separation of the clouds equal to about ten their
diameters. The dots are the results obtained by direct numerical
solution of the Gross-Pitaevskii equation (\ref{GPE_final}) and the
solid line is given by Eq.~(\ref{Delta_T_g_and_beta_zero}).
%
%
The optimized contrast of the interference fringes $V$ at the
recombination time $T + \Delta T$ for the parameters of
Fig.~\ref{fig:contrast_vs_T} is shown in
Fig.~\ref{fig:contrast_vs_T}. The dots correspond to the numerical
solution of the Gross-Pitaevskii equation and the solid line is
given by Eq.~(\ref{contrast_optimized}). For comparison, the dashed
line shows the contrast at the nominal recombination time $T$ given
by Eqs.~(\ref{contrast_simple}) and (\ref{deltak_recombination}). The lowest
values of the optimized contrast $V \approx 0.5$ correspond to
intermediate cycle times $T/R_{0} \approx 2$ when the maximum
separation between the two clouds is equal to their size. Both
increasing and decreasing the cycle time $T$ improves the contrast.

Figs.~\ref{fig:DeltaT_vs_T} and (\ref{fig:contrast_vs_T}) show that
the operation of the atom Michelson interferometer with the
optimization of the recombination time is possible both in the limit
$T/2R_{0} \le 1$ when the clouds overlap during all the cycle and in
the opposite limit $T/2R_{0} \gg 1$ when the clouds are separated
most of the time.

%
%
\subsection{Nonzero initial parabolic phase $g_{0} \ne 0$}
%
Performing an interferometric cycle with nonzero initial values of
the parabolic phase $g_{0}$ considerably improves the coherence as
compared to the case $g_{0} = 0$ provided the sign of $g_{0}$ is the
same as that of the nonlinearity $p$. The nonzero initial parabolic
phase can be acquired by relaxing the confinement frequency $\omega$
of the initial trap and letting the condensate evolve for some time
before the start of the interferometric cycle. Dynamics of the BEC
in time-dependent parabolic traps in Thomas-Fermi limit has been
extensively analyzed (see, e.g. \cite{castin96, dalfovo97,
kagan97}). In the case of 1D expansion corresponding to our
situation, the evolution of $g(\tau)$ and $r = R(\tau)/R(0)$ is
described by the set of equations
\begin{eqnarray}\label{BEC_expansion}
    &&\frac{d^{2}}{d\tau^{2}}r = - \omega^{2}(\tau)r +
    \frac{\omega^{2}(0)}{r^{2}}, \nonumber \\
    &&g = \frac{d}{d\tau}\ln r,
\end{eqnarray}
where $\omega(\tau)$ is the trap frequency. The exact value of $g$
depends on the detailed time dependence of $\omega(\tau)$. Changing
$\omega(\tau)$ adiabatically slowly leaves the phase of the
condensate flat, i.e. $g = 0$. Since we are interested in the
maximum possible value of $g$, we shall consider the limit when the
trap frequency is relaxed very fast so that $\omega(\tau) = 0$ for
$\tau > 0$. In this limit, $g$ is given by the relation (the
condensate's initial phase in the trap is zero):
\begin{equation}\label{g_expansion}
    g(r) = \left(\frac{3p}{R^{3}_{tr}}\right)^{1/2}\left(\frac{r -
    1}{r^{3}}\right)^{1/2},
\end{equation}
where $R_{tr}$ is the initial radius of the condensate in the trap.
An extra factor of two in Eq.~(\ref{g_expansion}) as compared to
Eq.~(\ref{g_longT}) is due to the fact that the initial condensate
is normalized to one whereas the two propagating clouds are
normalized to $1/2$.

For the given value of $R_{tp}$, g is maximum for $r = 3/2$. The
final size of the condensate after the expansion is the initial size
$R_{0}$ of the propagating clouds in the interferometric cycle,
i.e., $R_{0}/R_{tr} = 3/2$. The maximum possible value of $g_{0}$ is
thus given by the relation
\begin{equation}\label{g0_max}
    g_{0,max} = \left(\frac{3p}{2R^{3}_{0}}\right)^{1/2}.
\end{equation}
In the following we will use the value $g_{0} = sg_{0,max}$ where
the coefficient $0 \le s \le 1$ accounts for relaxing the trap with
finite speed.

\begin{figure} 
\includegraphics[width=8.6cm]{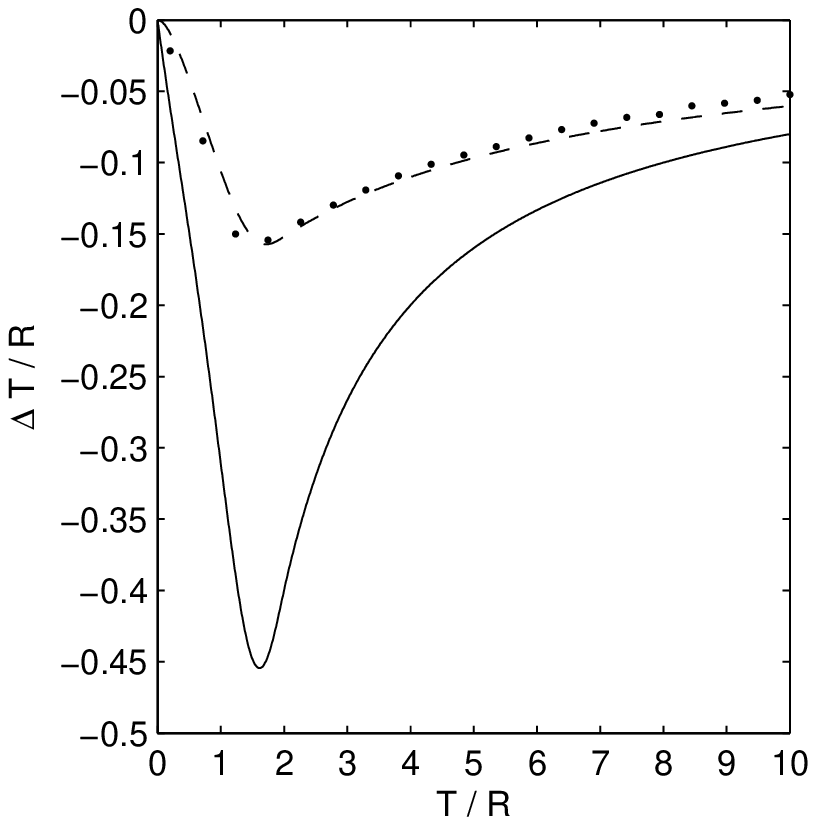}%
\caption{ \label{fig:DeltaT_vs_T_g0} The shift in the recombination
time $\Delta T/R$ as a function of the cycle time $T/R$ given by
Eq.~(\ref{Delta_T_nonzero_g0}) for $s = 0$ (solid) and $s = 0.2$
(dashed curve). The dots are the results of a numerical solution of
the GPE. For all three curves $R/p = 100$.}
\end{figure}

The correction to the recombination time $\Delta T/R$ given by the
equation (\ref{Delta_T_simple}) with $\beta = 0$ takes the form
\begin{equation}\label{Delta_T_nonzero_g0}
    \frac{\Delta T}{R} = \frac{- 2D_{1}(T/R)}{(3/4)(T/R) + D_{2}(T/R) + s(3R/2p)^{1/2}}
\end{equation}
If the parameter $(R/p)$ is large, which is typically the case, the
corrections to the recombination time are small and the contrast is
high. This is illustrated by Fig.~\ref{fig:DeltaT_vs_T_g0} which
shows the shift in the recombination time $\Delta T / R$ as a
function of the cycle time $T/R$ using
Eq.~(\ref{Delta_T_nonzero_g0}). The solid line corresponds to $s=0$
when the condensate does not have initial parabolic phase. The
dashed curve gives $\Delta T/R$ for $s = 0.2$, when the condensate
has been allowed to acquire initial parabolic phase. The dots are
the results of a numerical solution of the GPE with $s = 0.2$.
Figure \ref{fig:DeltaT_vs_T_g0} demonstrates that the shift in the
recombination time is considerably smaller when the condensate is
allowed to expand before the beginning of the cycle. Since the two
harmonics have larger overlap at the optimal recombination time, the
contrast in the interference fringes is larger when $g_0 \ne 0$.
%

\subsection{Nonzero parabolic external potential $\beta \ne 0$}
%
Nonzero values of the parabolic external potential $\beta \ne 0$ can
be due to environment or technical imperfections of an experimental
apparatus. Equation (\ref{Delta_T_simple}) with $\beta \ne 0$ and
$g_{0} = 0$ yields
\begin{equation}\label{Delta_T nonzero_beta}
    \frac{\Delta T}{R} = -
    \frac{1}{4}\frac{8D_{1}(\zeta) - (\beta R^{3}/p)\zeta^{2}}{3\zeta/4 + D_{2}(\zeta) - (\beta
    R^{3}/p)\zeta},
\end{equation}
where $\zeta = T/R$. The influence of the parabolic potential on the
operation of the atom Michelson interferometer is characterized by
the parameter $b = \beta R^{3}/p$. Note that since the term with
$\beta$ in the numerator of Eq.~(\ref{Delta_T nonzero_beta}) is
proportional to the square of the cycle time and the denominator
grows linearly with time, even small values of $b$ for long enough
cycles will always result in a complete loss of coherence.

In the limit of short cycle times $\zeta \ll 1$,
Eq.~(\ref{Delta_T_nonzero_g0}) takes the form
\begin{equation}\label{Delta_T_nonzero_beta_zeta_small}
    \frac{\Delta T}{R} = -\frac{\zeta}{4}\frac{3 - b}{5/2 - b}.
\end{equation}
Equation (\ref{Delta_T_nonzero_beta_zeta_small}) is similar to
Eq.~(\ref{Delta_T_g_and_beta_zero_zeta_small}) but the sign of
$\Delta T$ can be both negative and positive depending on the value
of $b$. The second difference is in that the coefficient multiplying
$\zeta$ may become so large for positive values of $b \approx 5/2$,
that coherence will be lost even for short cycle times. Negative
values of $\beta$ are preferable because they ensure the operation
of the interferometer at least for short times $T/R \le 1$. If the
value of $\beta$ is controlled at the level $b \ll 1$, the operation
of the interferometer is possible for $\Delta T/R < 1$ and any sign
of $\beta$.

In the limit $\zeta > 2$, Eq.~(\ref{Delta_T nonzero_beta}) becomes
\begin{equation}\label{Delta_T_nonzero_beta_zeta_large}
    \frac{\Delta T}{R} =  - \frac{12/5 - b\zeta^{2}}{(3 -
    4b)\zeta}.
\end{equation}
If $|b| \ll 1$, the optimized contrast will be high in the range
\begin{equation}
    2 \le \frac{T}{R} \ll \frac{1}{|b|}.
\end{equation}
If $|b| \ge 1$, the coherence in general will be lost for $T/R > 2$.

Additional limitations on the strength of the quadratic potential
$\beta$ and the cycle time $T$ are due to the fact that the
reflection pulses will not operate well if the relative change in
the velocity of the atomic cloud $|\Delta v|/v_{0}$ exceeds about
$0.1$. Using the fact that the nominal dimensionless velocity $v_{0}
= 1$ and the relation $|\Delta v| = (1/2)|\beta|(T/2)^{2}$, one gets
$|\beta| T^{2} \le 0.8$.

%
\subsection{Recombination at a different wavelength}
%
The contrast of the interference fringes can be improved by
conducting the recombination with optical pulses having different
wavelength as compared to the splitting pulse to compensate for the
change in the wave vectors of the moving clouds. The relative change
in the wavelength of the recombining pulse $\Delta \lambda/\lambda$
as compared to the separation pulse is given by the expression (cf.
Eq.~(\ref{v_Final}))
\begin{equation}\label{Delta_lambda}
    \frac{\Delta \lambda}{\lambda} = -\Delta \kappa = -
    \frac{\beta}{2}T^{2} + \frac{4p}{R}D_{1}(T/R).
\end{equation}
As has been discussed in the introduction, the repulsive
nonlinearity results in the speeds $v$ of the moving harmonics
$\psi_{\pm}$ being smaller than the speed $v_{0}$ imparted by the
separation pulse. The recombination then should be performed with
beams of larger wavelength. Similarly, for $\beta < 0$ (a potential
hump) $\Delta \lambda > 0$ and for $\beta > 0$ (a potential trough)
$\Delta \lambda < 0$. The optimized contrast is determined by the
relation
\begin{equation}\label{contrast_Delta_lambda}
    V \approx 1 - \frac{3}{2}\left(\frac{\Delta x}{2R}\right)^{2}\left[\ln
    \frac{2R}{|\Delta x|} + 2\ln2 - \frac{1}{2}\right],
\end{equation}
where $\Delta x$ is the separation between the centers of the
harmonics $\psi_{\pm}$ at the recombination time given by
Eq.~(\ref{x_Final}). In a typical situation, $|\Delta x|/R \ll 1$
and the optimized contrast is close to one.

\section{Acknowledgements}
This work was supported by the Defense Advanced Research Projects
Agency (Grant No. W911NF-04-1-0043).



\appendix
\section{Dynamics of the BEC due to the optical pulses}
The optical potential is used to split the initial zero-momentum BEC
cloud at the beginning of the interferometric cycle into the two
harmonics with the momenta $\pm 1$, reverse their direction of
propagation in the middle of the cycle and recombine them at the
end. The optical pulses are short and sufficiently intense so that
the dynamics of the condensate is dominated by the optical potential
when the laser beams are on and the diffraction, relative
displacements of the clouds and the nonlinearity can be neglected. A
good quantitative description of the BEC dynamics can be obtained
keeping only the lowest three harmonics with $n = 0, \pm 1$ in
Eq.~(\ref{psi_n_harmonics}). The set of
Eq.~(\ref{eqns_for_harmonics}) with these approximations reduces to
\begin{equation}\label{eqn_43harm2}
    i\frac{d}{d\tau}\left[
    \begin{array}{c}
    \psi_{-1} \\
    \psi_{0} \\
    \psi_{1} \end{array}\right]
    = \frac{1}{2}\left[
    \begin{array}{ccc}
    1 & \Omega & 0 \\
    \Omega & 0 & \Omega \\
    0 & \Omega & 1
    \end{array}\right]
    \left[
    \begin{array}{c}
    \psi_{-1} \\
    \psi_{0} \\
    \psi_{1} \end{array}\right].
\end{equation}
Solution of Eq.~(\ref{eqn_43harm2}) has the form
\begin{equation}\label{matrixA}
    \left[\begin{array}{c}
    \psi_{-1}(\tau) \\
    \psi_{0}(\tau) \\
    \psi_{1}(\tau)
          \end{array}
\right]
 = \left[\begin{array}{ccc}
    A_{11} & A_{12} & A_{13} \\
    A_{12} & A_{22} & A_{12} \\
    A_{13} & A_{12} & A_{11}
    \end{array}\right]
    \left[\begin{array}{c}
    \psi_{-1}(0) \\
    \psi_{0}(0) \\
    \psi_{1}(0)
          \end{array}
\right]
\end{equation}
where
\begin{eqnarray}\label{matrixAelements}
    A_{11} &=& \frac{1}{2}\left[\cos \frac{s\tau}{4} + e^{-i\tau/4}
    -\frac{i}{s}\sin \frac{s\tau}{4}\right], \\
    A_{12} &=& -2i\frac{\Omega}{s}\sin \frac{s \tau}{4} , \\
    A_{13} &=& \frac{1}{2}\left[\cos \frac{s\tau}{4} - e^{-i\tau/4}
    - \frac{i}{s}\sin \frac{s\tau}{4}\right], \\
    A_{22} &=& \cos \frac{s \tau}{4} +
    \frac{i}{s}\sin \frac{s\tau}{4},
\end{eqnarray}
and $s = \sqrt{1 + 8 \Omega^2}$.
Using Eq.~(\ref{matrixA}) it is straightforward to show that the
momentum reversal of the moving BEC clouds $\psi_{\pm 1} \rightarrow
\psi_{\mp 1}$ can be achieved with a single pulse of duration
$\tau_p = 4 \pi$ and magnitude $\Omega_{p} = (3/8)^{1/2}$. The
unitary evolution matrix corresponding to the momentum reversal
pulse is of the form
\begin{equation}
    U_{\pm 1 \leftrightarrow \mp 1} =
    \left[\begin{array}{ccc}
    0 & 0 & 1 \\
    0 & 1 & 0 \\
    1 & 0 & 0
    \end{array}\right].
\end{equation}
Splitting of the zero-momentum cloud $\psi_{0}$ into the two
harmonics $\psi_{\pm 1}$ and the recombination (the inverse of the
splitting) requires a double pulse sequence. The first pulse with
$\Omega_p = (1/8)^{1/2}$ and $\tau_p = 2^{1/2} \pi$ is followed by a
period of free evolution when the lasers are turned off for a time
interval $\tau_{ev} = 2 \pi$ and then followed by the second optical
pulse with $\Omega_p = (1/8)^{1/2}$ and $\tau_p = 2^{1/2} \pi$. The
evolution matrix for the splitting sequence is given by
\begin{equation}\label{matrix_2pulse_optimized}
    U_{0 \leftrightarrow \pm 1} = \left[\begin{array}{ccc}
    -\frac{1}{2}\exp(-i\pi/\sqrt{2}) & \frac{1}{\sqrt{2}} &
    \frac{1}{2}\exp(-i\pi/\sqrt{2})\\
    \frac{1}{\sqrt{2}} & 0 & \frac{1}{\sqrt{2}} \\
    \frac{1}{2}\exp(-i\pi/\sqrt{2}) & \frac{1}{\sqrt{2}} &
    - \frac{1}{2}\exp(-i\pi/\sqrt{2})
    \end{array}\right],
\end{equation}
(irrelevant common phase has been omitted).



\end{document}